\documentclass[sigconf]{acmart}
\acmConference[ICSE 2022]{The 44th International Conference on Software Engineering}{May 21-29, 2022}{Pittsburgh, PA, USA}

\AtBeginDocument{%
  \providecommand\BibTeX{{%
    \normalfont B\kern-0.5em{\scshape i\kern-0.25em b}\kern-0.8em\TeX}}}

\setcopyright{acmcopyright}
\copyrightyear{2022}
\acmYear{2022}
\usepackage{amsmath,amsfonts}
\usepackage{algorithmic}
\usepackage{graphicx}
\usepackage{textcomp}
\usepackage{xcolor}
\usepackage{xspace}
\usepackage{booktabs}
\usepackage{multirow}
\usepackage{subfig}
\usepackage{verbatimbox}
\usepackage{fancyvrb, listings}
\usepackage{courier}
\usepackage[capitalise,noabbrev]{cleveref}
\newcommand\lt[1]{{\lstinline[style=cstyle4]!#1!}}
\lstdefinestyle{cstyle4}{
language=c,
basicstyle=\ttfamily\bfseries\footnotesize,
  morekeywords={virtualinvoke},
  keywordstyle=\color{blue},
  ndkeywords={result},
  ndkeywordstyle=\color{red},
  numbers=none,
  breaklines=true,
  numbersep=10pt,
  backgroundcolor=\color{white},
  tabsize=2,
  showspaces=false,
  showstringspaces=false,
  xleftmargin=0,
  captionpos=b,
  escapeinside={$}{$},
  print
}

\definecolor{dkgreen}{rgb}{0,0.5,0}

\newcommand{\modelname}{{\sc VarCLR}\xspace}
\newcommand{\modelavg}{{\sc VarCLR-Avg}\xspace}
\newcommand{\modellstm}{{\sc VarCLR-LSTM}\xspace}
\newcommand{\modelbert}{{\sc VarCLR-CodeBERT}\xspace}
\newcommand{\wordtvec}{\texttt{word2vec}\xspace}
\newcommand{\dataname}{{\sc GitHubRenames}\xspace}
\newcommand{\idbench}{{\sc IdBench}\xspace}
\newcommand{\idsim}{{\sc VarSim}\xspace}
\newcommand{\vartypo}{{\sc VarTypo}\xspace}
\newcommand{\red}[1]{\textcolor{red}{#1}}

\newcommand{\hide}[1]{}

\begin{document}

% \title{Claire: A Contrastive Learning Approach of Identifier REpresentation\\
% \title{COVID’r: Contrastive learning Of Variable/IDentifier representation\\
\title{VarCLR: Variable Semantic Representation Pre-training via Contrastive Learning}

\author{Qibin Chen}
\email{qibinc@cs.cmu.edu}
\affiliation{
    \institution{Carnegie Mellon University}
    \country{}
}

\author{Jeremy Lacomis}
\email{jlacomis@cs.cmu.edu}
\affiliation{
    \institution{Carnegie Mellon University}
    \country{}
}

\author{Edward J. Schwartz}
\email{eschwartz@cert.org}
\affiliation{
    \institution{Carnegie Mellon University Software Engineering Institute}
    \country{}
}

\author{Graham Neubig}
\email{gneubig@cs.cmu.edu}
\affiliation{
    \institution{Carnegie Mellon University}
    \country{}
}

\author{Bogdan Vasilescu}
\email{bogdanv@cs.cmu.edu}
\affiliation{
    \institution{Carnegie Mellon University}
    \country{}
}

\author{Claire Le~Goues}
\email{clegoues@cs.cmu.edu}
\affiliation{
    \institution{Carnegie Mellon University}
    \country{}
}

\begin{abstract}
    % CLG is going back and forth on that...
    Variable names are critical for conveying intended program behavior.
    Machine learning-based program analysis methods use variable name representations
    for a wide range of tasks,
    such as suggesting new variable names and bug detection.
    Ideally, such methods could capture semantic relationships between
    names beyond syntactic similarity, e.g., the fact that the
    names \lt{average} and \lt{mean} are similar.
    % Previous methods adopt embedding techniques based on
    % the distributional hypothesis
    % to produce representations intended to capture such relationships.
    Unfortunately, previous work has found that even the best of previous representation approaches primarily capture
    ``relatedness'' (whether two variables are linked at all),
    rather than ``similarity'' (whether they actually have the same meaning).

    We propose \modelname, a new approach for learning semantic representations
    of variable names that effectively captures variable similarity in this
    stricter sense.
    We observe that this problem is an excellent fit for
    \emph{contrastive learning}, which aims to minimize the distance between explicitly similar
    inputs, while maximizing the distance between dissimilar
    inputs. This requires labeled training data, and thus we construct a novel,
    weakly-supervised variable renaming dataset mined from GitHub
    edits.
    We show that \modelname enables the effective application of
    sophisticated, general-purpose language models like BERT,
    to variable name representation and thus also to
    related downstream tasks like variable name similarity search or spelling correction.
    \modelname produces models that
    significantly outperform the state-of-the-art
    on \idbench, an existing benchmark that explicitly
    captures variable similarity (as distinct from relatedness).
    Finally, we contribute a release of all data, code, and pre-trained models,
    aiming to provide a drop-in replacement for  variable representations
    used in either existing or future program analyses that rely on
    variable names.
    %    \modelname surpasses the state-of-the-art with 20.3\% and 14.7\%
    %    absolute improvement in Spearman's rank correlation over the
    %    previous best method \es{should we say what this is?} on the recent
    %    IdBench-small and IdBench-large benchmarks.  
\end{abstract}

\maketitle

\section{Introduction}
\label{sec:intro}

% maybe note somewhere that IDBench is a call to arms on this, like, to make
% things better.

Variable names convey key information about code
structure and developer intention. They are thus central for code comprehension, readability, and
maintainability~\cite{Lawrie2006,binkley2013impact}. A growing array of
automatic techniques make use of variable names in the context of tasks
like but not limited to bug finding~\cite{pradel2011detecting,rice2017detecting}
or specification mining~\cite{zhong09}. Beyond leveraging the information
provided by names in
automated tools, recent work has increasingly attempted to directly suggest good
or improved names, such as in reverse
engineering~\cite{Jaffe2018,lacomis2019dire} or
refactoring~\cite{Allamanis2014,Context2Name,liu2019learning}.

Developing (and evaluating) such automated techniques (or \emph{name-based
    analyses}~\cite{wainakh2021idbench}) relies in large part on the ability to model and
reason about the relationships between variable names. For concreteness,
consider an analysis for automatically suggesting names in decompiled
code. Given a compiled program (such that variable names are
discarded) that is then decompiled (resulting in generic names like
\lt{a1},\lt{a2}), a renaming tool seeks to replace the generic
decompiler-provided identifiers with more informative variable names for the
benefit of reverse engineers aiming to understand it. Good names in this context
are presumably closely related to the names used in the original program (before
the developer-provided names were discarded). A variable originally named
\lt{max}, for example, and then decompiled to \lt{a2}, should be
replaced with a name at least close to \lt{max}, like \lt{maximum}.
Modeling this relationship well is key for both constructing and evaluating such analyses.

Accurately capturing and modeling these relationships is difficult.  A
longstanding approach has used syntactic difference --- like various
measures of string edit distance --- to estimate the relationship between two
variables (such as for spellchecking~\cite{spellcheck}). However,
syntactic distance is quite limited in capturing underlying name semantics.
For example, the pairs (\lt{minimum}, \lt{maximum}) and
(\lt{minimum}, \lt{minimal}) are equidistant syntactically --- with  a Levenshtein
distance of two --- but \lt{maximum} and \lt{minimum} are antonyms.

More recent work has sought to instead encode variable name semantics using neural
network embeddings, informing a variety of name-based
analyses~\cite{pradel2018deepbugs,harer2018automated,white2019sorting}.
Unfortunately, although state-of-the-art techniques for
variable name representation better capture
\emph{relatedness}, they still
struggle to accurately capture variable name
\emph{similarity}, in terms of how interchangeable two names are.
Variables may be related
for a variety of reasons.  While \lt{maximum} and
\lt{minimum} are highly related, they certainly cannot
be substituted for one another in a
code base. \lt{minimum} and \lt{minimal}, on the other hand, are both
related and very similar.  In recent work, Wainakh et
al.~\cite{wainakh2021idbench} presented a novel dataset, \idbench,
based on a human survey on variable similarity and
interchangeability, and used it to evaluate state-of-the-art embedding approaches.
They empirically established that there remains significant room for
improvement in terms of capturing similarity rather than merely
relatedness.
%
%Even
%the best embedding model which is based on FastText~\cite{bojanowski2017enriching} has enormous margins
% the best ensemble learning approach most recently proposed has enormous margins
%% Note: their results of the ensembling learning approach are actually quite good,
%% since it trained on IdBench. so for each sample, 
%% they trained a SVM classifier on all other samples in idbenchand predict for this sample
%% this is usually used as leave-on-out cross validation and *not legit*
%% for submitting to a benchmark --QC
%% The above is explained in Sec 3.2.1 Baselines
%for improvement on this key measure~\cite{wainakh2021idbench}. 

In this paper, we formulate the \emph{variable semantic representation learning
    problem} as follows: given a set of variable data, learn a function
$f$ that maps a variable name string to a low-dimensional dense vector that can
be used in a variety of tasks (like the types of
name-based analyses discussed above).  To be useful, such a mapping function
should effectively encode \emph{similarity}, i.e., whether two variables have the same
meaning.  That is, $f(\mathtt{minimum})$ and   $f(\mathtt{minimal})$ should be
close to one another.  Importantly, however, the function
should also ensure that variable names that are \emph{not} similar (regardless
of relatedness!) are \emph{far} from one another.  That
is, $f(\mathtt{minimum})$ and $f(\mathtt{maximum})$ should be distant.

Our first key insight is that this problem is well suited for a
contrastive-learning approach~\cite{wu2018unsupervised,oord2018representation,he2020momentum,chen2021exploring}.
% {tian2020contrastive,chen2020simple,caron2020unsupervised,grill2020bootstrap}
Conceptually, contrastive
learning employs encoder networks to encode instances (in this task, variables)
into representations (i.e., hidden vectors), with a goal of minimizing the
distance between (the representation of) similar instances and maximizing the
distance between (the representation of) dissimilar instances.
Contrastive learning requires as input a set of ``positive pair''
examples---of similar variables, in our case---for training.

Our second key insight is that we can construct a suitable weakly-supervised
dataset of examples of similar variables by taking advantage of large
amounts of source control information on GitHub.  Following the definition of
``similarity'' from prior work~\cite{miller1991contextual,wainakh2021idbench},
we consider two variable names are similar if they have the same meaning, or are
\emph{interchangeable}.  We therefore automatically mine source control
edits to identify historical
changes where developers renamed a variable but did not otherwise overly
modify the code in which it was used.  Although potentially noisy, this
technique matches an
intuitive understanding of variable name similarity in terms of
interchangeability, and allows for the collection of a large dataset, which we
call \dataname.

Finally, we observe that the variable semantic representation learning problem
requires more powerful neural architectures
than \texttt{word2vec}-based
approaches~\cite{mikolov2013distributed,bojanowski2017enriching,wainakh2021idbench}.\footnote{word2vec~\cite{mikolov2013distributed} is an embedding algorithm based on the
    \emph{distributional hypothesis}, which assumes
    words that occur in the same contexts tend to have similar meanings.}
Such approaches are limited both empirically (as Wainakh et al. showed) and
conceptually; note for example that  they cannot capture component ordering,
such as the difference between \lt{idx_to_word} and \lt{word_to_idx}.
Meanwhile, Pre-trained Language Models (PLMs)~\cite{devlin2019bert,raffel2020exploring,brown2020language}
based on the powerful Transformer architecture~\cite{transformer}
have achieved the state-of-the-art on a wide range of natural language processing tasks,
including text classification~\cite{devlin2019bert},
question answering and summarization~\cite{lewis2020bart},
and dialog systems~\cite{adiwardana2020towards}.
PLMs tailored specifically for programming languages
such as CodeBERT~\cite{feng2020codebert} and Codex~\cite{chen2021evaluating}
are useful in a variety of tasks such as code completion,
repair, and generation~\cite{lu2021codexglue,chen2021evaluating}, though not
yet for variable name representation.
Encouragingly, previous work shows that contrastive learning can strongly
improve BERT sentence embeddings for textual similarity
tasks~\cite{gao2021simcse}. And,
contrastive learning
has been shown to benefit from deeper and wider network architectures~\cite{chen2020simple}.

We combine these insights to produce \modelname, a novel machine learning
method
based on contrastive learning for learning general-purpose variable
semantic representation encoders. In \modelname, the contrastive learning
element serves as a pre-training step for a traditional encoder.
While powerful modern approaches like CodeBERT perform poorly on the variable
representation problem off-the-shelf, we show that \modelname-trained
models dramatically outperform the previous state-of-the-art on capturing both
variable similarity and relatedness.
\modelname is designed to
be general to a variety of useful downstream tasks; we
demonstrate its effectiveness for both the basic variable similarity/relatedness
task (using the
\idbench dataset as a gold standard baseline) as well as for variable similarity
search, and spelling error correction.

To summarize, our main contributions are as follows:

\begin{enumerate}
    \item \modelname, a novel method based on contrastive learning that learns
          general-purpose variable semantic representations suitable for a
          variety of downstream tasks.
    \item A new weakly supervised dataset, \dataname,
          for better variable representation learning consisting of similar
          variable names collected from real-world GitHub data.
    \item Experimental results demonstrating that \modelname's models
          significantly outperform state-of-the-art representation approaches on \idbench,
          an existing benchmark for evaluating variable semantic
          representations.
          These results further substantiate the utility of more
          sophisticated models like CodeBERT,
          with larger model capacity, in place of the previous
          \texttt{word2vec}-based methods for learning variable representations,
          while showing that the contrastive learning pre-training step is
          critical to enabling the effectiveness of such models.
    \item Experimental results that demonstrate that both unsupervised
          pre-training and our proposed
          weakly-supervised contrastive pre-training are indispensable parts for
          advancing towards the state-of-the-art,
          for the former takes advantage of greater \textit{data quantity}
          by leveraging a huge amount of unlabeled data,
          while the latter takes advantage of better \textit{data quality}
          with our new \dataname dataset.
\end{enumerate}

Finally, we contribute a release of all data, code, and pre-trained models,
aiming to provide a drop-in replacement for  variable representations
used in either existing or future program analyses that rely on
variable names.%
\footnote{Code, data, and pre-trained model available at \url{https://github.com/squaresLab/VarCLR}.}
% Quality: with respect to github generally rather than with respect to idbench. 

% How do we compare to the previous models? Studied effect of training data on our
% own models. 

\section{Problem Domain}
\label{sec:problem}

Variable names critically communicate developer intent and are thus increasingly
used by a variety of automated techniques as a central source of information.
Such techniques
increasingly rely on machine learning and embedding-based representation
approaches to encode variable name meaning for these purposes. However, recent
work~\cite{wainakh2021idbench}
shows that while neural embeddings based on techniques like
\texttt{word2vec} do a better job of capturing relationships between variables
than syntactic edit distance does, they still struggle to capture actual variable
similarity in terms of their interchangeability.
In this paper, we show that this problem is amenable to a contrastive learning
approach, enabling accurate general-purpose representations of variable name
semantics.

We define the \textit{variable semantic representation learning problem} as
follows: given a collection of suitable variable data, learn a function $f$ that
maps a variable name string to a low-dimensional dense vector that can be used
to benefit various downstream tasks (like variable similarity scoring in the
simplest case, or arbitrarily complex name-based analyses).
A good mapping function $f$ for variable name representations
should:
\begin{enumerate}
    \item \emph{Capture similarity.}
          % Justify our use of \dataname dataset instead of unsupervised word2vec methods in IdBench
          $f$ should encode \emph{similar} names such that
          they are close to one another.  Two names are similar when they have
          similar or generally interchangeable meanings, like \lt{avg} and
          \lt{mean}.
          This is especially important for variables that
          are \emph{related} but \emph{not similar}, such as \lt{maximum} and
          \lt{minimum}.  Indeed, antonyms are often closely
          related and can appear in similar contexts (\lt{max} and \lt{min} for
          example may be used together in loops finding extrema).
    \item \textit{Capture component ordering and importance.}
          % Justify our use of LSTM/BERT compared to word embedding averaging in IdBench
          Variables often consist of component words or sub-words.  We observe
          that the order of such components can affect meaning.
          For example, \lt{idx\_to\_word} and
          \lt{word\_to\_idx} contain the same sub-words, but have different
          meanings.
          Moreover, the importance of different component words in a variable
          can be different and the importance of the same word can vary between
          variables.
          For example, in variables \lt{onAdd} and \lt{onRemove},
          \lt{on} is less important, while \lt{add} and \lt{remove} are more
          important. In \lt{turnOn} and \lt{turnOff},
          \lt{on} and \lt{off} are more important than \lt{turn}.
          A good mapping function $f$ should be able to
          capture these differences, instead of treating variables as an unordered
          bag of sub-words.
    \item \textit{Transferability.}
          % Justify our use of contrastive learning framework
          % as general-purpose pre-training instead of focusing on a specific task
          The representation should be general-purpose and usable for a wide range of tasks.
          Benefits of a transferable, shared representation include the ability to (1) improve accuracy on unsupervised or data-scarce tasks, where it can be hard to
          obtain high-quality variable representations from scratch, and (2) for
          complex tasks consisting of many sub-tasks,
          make better use of labeled data from multiple sub-tasks via multi-task
          learning.
          % \gn{I commented this out as it didn't seem necessary.}
          % General pre-trained representations that
          % correlates with human understanding better than heuristics
          % can provide a better initialization for such tasks.
\end{enumerate}

% \es{Qibin please make sure this paragraph is not BS.}
This formulation of the problem motivates our use of \emph{contrastive
    learning}, which is an effective way to learn similarity from
labeled data. %  Conceptually, contrastive learning aims to minimize the distance
% between the representations of positive pairs, while maximizing the distance
% between negative pair examples. 
Conceptually, given an encoder network $f_\theta$ and a set of
% labeled  %% note: positive pairs can be created from unlabeled data with aug
similar ``positive pairs'', contrastive learning returns a
\emph{new} encoder that attempts to locate similar ``positive pair''
instances closer together and dissimilar ``negative pair'' instances
farther apart.
In practice, this can be accomplished by re-training the original encoder on a new
pre-training
% unsupervised  %% note: could be unsupervised/supervised according to how the similar pairs are created
task: instance
discrimination~\cite{wu2018unsupervised}.
Instance discrimination casts the contrastive learning problem as a
classification problem where only the ``positive pair'' instances are
equivalent.  Rather than explicitly adjusting the distances between
points, the encoder's parameters are trained to optimize its
performance at discriminating similar instance from dissimilar
instances.  This naturally adjusts the parameters of the encoder such
that similar instances are moved closer together (and vice-versa for
dissimilar instances).
The actual output of the contrastive learning process is a new encoder
$f_{\theta'}$ that is identical to the original encoder in neural architecture,
but has a different set of parameters $\theta'$ resulting from
training on the instance discrimination task.

% This formulation of the problem motivates our use of \emph{contrastive
%   learning}, which is an effective way to learn similarity from labeled data.
% Conceptually, contrastive learning is based on a simple high-level idea ---
% pulling close the encoded representations of ``positive pair'' instances,
% while pushing away the representations of ``negative pair'' instances.

There are two central design choices in applying contrastive learning, however.
First, \emph{Which neural architectures should be used for $f_\theta$?} This is
usually decided by the problem domain in question. For example, in computer
vision, ResNet~\cite{he2016deep} for learning image
representations~\cite{oord2018representation,he2020momentum,chen2020simple}; in
natural language processing, Simple word embedding or
BERT~\cite{devlin2019bert,liu2019roberta} for learning sentence
representations~\cite{wieting2019simple,gao2021simcse}; and in data mining,
Graph Neural Network~\cite{kipf2017semi} for learning graph
representations~\cite{qiu2020gcc}. Second,
\emph{How to construct similar (positive) and dissimilar (negative) training pairs?}
Unsupervised data augmentation like cropping or
clipping has been used to create different ``views'' of the same image as
similar pairs in image
processing~\cite{oord2018representation,tian2020contrastive}; word dropout can
augment text sentences for natural language processing~\cite{gao2021simcse}. For
supervised contrastive learning, positive pairs
can be created from labeled datasets directly~\cite{gao2021simcse}, or via
sampling instances from the same class~\cite{khosla2020supervised}. Note that
dissimilar pairs typically need not be explicitly defined.
Instead, \textit{in-batch
    negatives}~\cite{oord2018representation} can be sampled from instance pairs
that are not explicitly labeled as positive.

The choice of similar instances is very important,
as it influences the learned similarity function
and impacts downstream effectiveness~\cite{tian2020makes}.
For example, consider how training can lead to unintentional properties of a learned similarity
function for \texttt{word2vec}.
At a high level, \wordtvec~\cite{mikolov2013distributed} can be viewed as a form of
unsupervised contrastive learning.
It employs a word embedding layer as the encoder,
and treats words co-occurring in the context window as similar pairs,
while treating other words in the dictionary as dissimilar ones.\footnote{
    We leave out the minor difference that \wordtvec produces two sets of embeddings,
    while contrastive learning usually uses a unified representation.}
Due to its choice of ``similar instances'', it learns more of association (or relatedness) between words,
instead of similarity in terms of how interchangeable two words are.
For example, \wordtvec embeddings of cohyponym words such as red, blue, white, green are very close.
While this might not be a problem in NLP applications,
\wordtvec leads to unsatisfactory behavior when applied to variable names~\cite{wainakh2021idbench},
e.g., by identifying \lt{minLength} and \lt{maxLength} as similar.

\section{Method}
\label{sec:method}

\begin{figure}
    \centering
    \includegraphics[width=\linewidth]{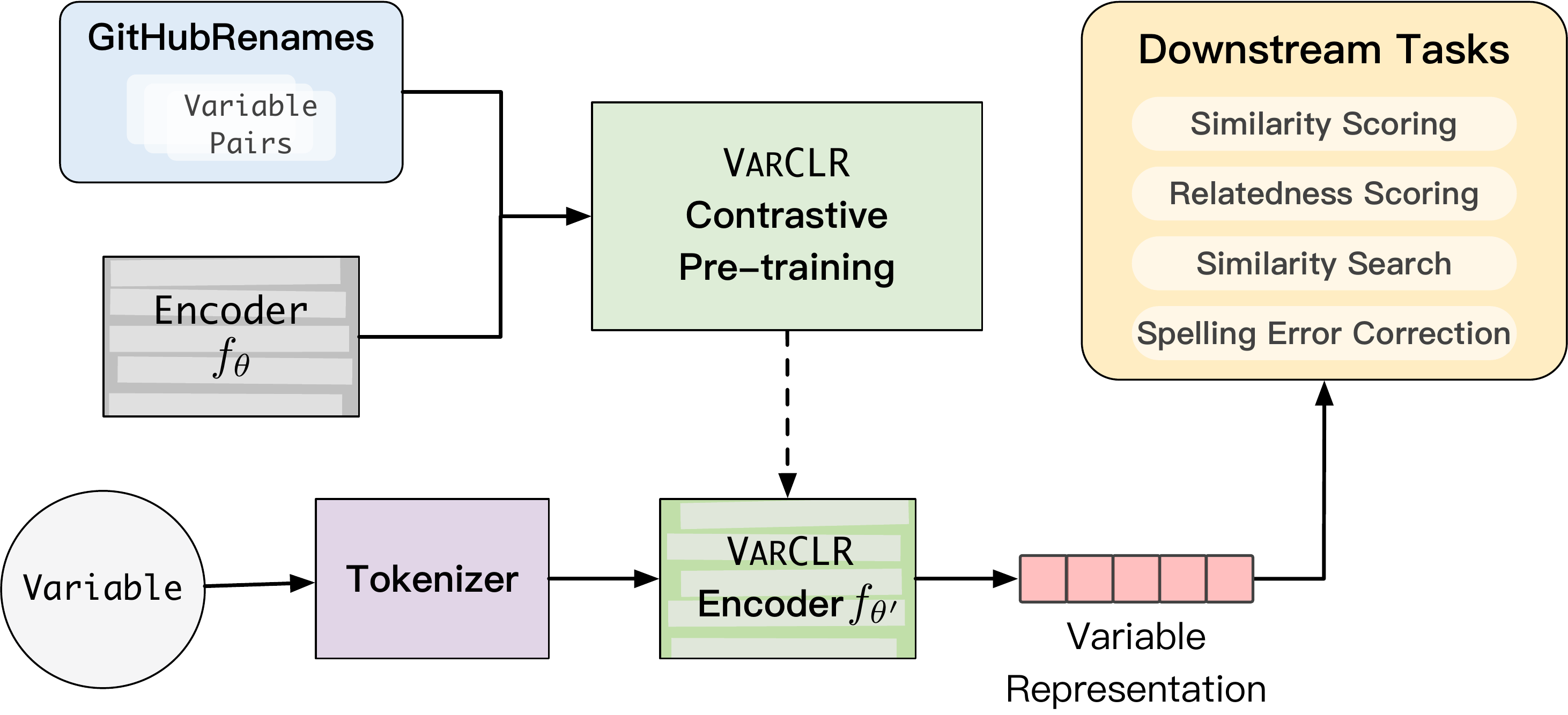}
    \caption{\small Conceptual overview of \modelname.
        %\gn{I don't know if I'd call this an ``architecture''? Maybe ``Conceptual overview of \modelname''}
    }
    \label{fig:arch}
\end{figure}

\cref{fig:arch} shows a high-level conceptual overview of \modelname,
our framework for learning effective semantic representations of
variable names.  \modelname consists of a contrastive pre-training
phase that
takes two inputs: (1) a positive set of similar variable name pairs, and (2) an
input encoder.
The set of similar variables is crucial for
\modelname's performance. We thus produce \dataname, a novel
weakly-supervised dataset consisting of positive examples of similar
variables by examining large amounts of source code history
available from GitHub (Section~\ref{sec:dataset}).
These variables must be suitably
tokenized for encoding in a way that captures and retains relevant information
(Section~\ref{sec:tokenization}), both for pre-training and for downstream tasks.
\modelname also takes an input encoder $f_\theta$ with learnable
parameters $\theta$ (Section~\ref{sec:encoders}).  This encoder is then trained using contrastive learning
(Section~\ref{sec:pretraining}).
% We evaluate the performance impact
% of different input encoders in Section~\ref{sec:scoring}, and find
% that \modelname performs best using CodeBERT as an input
% encoder.\es{Qibin please sanity check me}
% CLG says: that's true but perhaps doesn't need to be mentioned up here yet?
%
The output of our framework is a contrastively-trained \modelname encoder
that converts tokenized variables into semantic
representations suitable for a variety of tasks and name-based analyses, including
similarity scoring or spelling error correction, among others.

\subsection{Similar variables: \dataname}
\label{sec:dataset}
%FIXME
% As an example of a ``positive pair'', in computer vision,
% unsupervised data augmentation approaches (e.g., cropping, horizontal clipping) can be used
% to create multiple ``views''~\cite{tian2020contrastive} (similar instances) of
% the same images.  

A high-level definition of ``similarity''~\cite{miller1991contextual,wainakh2021idbench},
is the degree to which two variables have the same meaning.
Contrastive learning requires positive examples for training, and thus we need
a set of appropriate positive pairs of similar variable names. As discussed in
Section~\ref{sec:problem}, these need not be manually constructed.
Although \idbench~\cite{wainakh2021idbench} provides curated sets of human-judged
``similar'' variables, they are too small for training
purposes (the largest set,
has 291 variable pairs). This motivates an automated mechanism for
constructing training data, with the added benefit that we need not be concerned
about training and testing on the same dataset (as we use \idbench for evaluation).

Instead, we observe that one way to define variable similarity is
to consider the degree to which two variables are explicitly
\emph{interchangeable} in code (close to \idbench's definition of
``Contextual similarity'').  We therefore collect a weakly supervised dataset of
interchangeable variable names by mining source control version histories for
commits where variable names change.  These variable pairs are
considered similar because they appear interchangeable in the same code
context.

Concretely, we built upon existing open-source dataset collection code used to mine source
control for the purpose of modeling
changes~\cite{yin2018learning}.\footnote{\url{https://github.com/microsoft/msrc-dpu-learning-to-represent-edits}}
Given a repository, this code mines all commits of less than six lines of
code where a variable is renamed.
The intuition is to look for commits that do not make large structural changes
that might correspond to a major change in a variable's meaning.
We applied dataset collection to an expanded version of the list of repositories used in
ref~\cite{yin2018learning}, consisting of 568 C\# projects.\footnote{\url{https://github.com/quozd/awesome-dotnet}}
The final \dataname dataset contains 66,855 variable pairs,
each consisting of a variable name before and after a renaming commit.

The \dataname dataset is only weakly supervised since developers were not asked
to label variable pairs explicitly.  The dataset may thus be noisy, and in
particular we did not attempt to filter out renames corresponding to bug fixes.
Indeed, we note that a number of pairs
in \dataname correspond to fixing spelling mistakes
(Section~\ref{sec:spelling}).
Overall, however, we note that our method transfers well to the \idbench
validation set, and expect that more data will only improve \modelname's effectiveness.

\subsection{Input representation}
\label{sec:tokenization}

A variable name as a text string must be preprocessed to be used as input to a neural network encoder.
We observe two interesting aspects of variable names that inform our preprocessing.
First, variable names are often composed of multiple words with interchangeable case styles,
e.g., \lt{max\_iteration} vs \lt{maxIteration}.
Second, variable names are sometimes composed of short words or abbreviations,
without an underscore or uppercase to separate them.
e.g., \lt{filelist}, \lt{sendmsg}.

For the first problem, we apply a set of regex rules to canonicalize variable names
into a list of \textit{tokens}, e.g., \lt{["max", "iteration"]}.
The second problem is more challenging,
and could cause Out-of-vocabulary (OOV) problems.
To solve this, we use the pre-trained CodeBERT tokenizer~\cite{feng2020codebert},
which is underlying a Byte Pair Encoding (BPE) model~\cite{sennrich2016neural}
trained on a large code corpus based on token frequencies.
When encountering an unknown composite variable name such as \lt{sendmsg},
it is able to split it into \textit{subword tokens}, e.g., \lt{["send", "\#\#msg"]},
where \lt{"\#\#"} means this token is a suffix of the previous word.

\subsection{Encoders}
\label{sec:encoders}

Generally, a neural encoder takes the input sequence, and
encodes and aggregates information over the sequence to produce a hidden vector.
That is,
given a sequence of tokens $v = (v_1, v_2, \ldots, v_n)$ corresponding to a
tokenized variable name,
an encoder outputs a hidden vector $\boldsymbol{h} \in \mathcal{R}^d$,
where $d$ is the dimension of the hidden representation:
\begin{align}
    \boldsymbol{h} = f_\theta(v),
\end{align}
$f_\theta$ denotes the encoder with learnable parameters $\theta$.

Note that \modelname is applicable to any
encoder with this form.
In this paper, we instantiate it specifically
for Word Embedding Averaging (\modelavg), the
LSTM Encoder (\modellstm), and BERT (\modelbert).

\vspace{-1ex}
\paragraph{Word Embedding Averaging}
Averaging the embeddings of input tokens is
a simple but effective way to represent a whole input sequence, given sufficient
data~\cite{wieting2019simple,wieting2021paraphrastic}.
Therefore, we consider this as a simple baseline encoder.
Formally, given the tokenized variable name
$v = (v_1, v_2, \ldots, v_n), v_i \in \mathcal{V}$,
and a word embedding lookup table $L_\theta: \mathcal{V} \rightarrow \mathcal{R}^d$:
\begin{align}
    \boldsymbol{h} = \frac{1}{n}\sum_{i=1}^{n}L_{\theta}(v_i),
    \label{eq:word_avg}
\end{align}
where $\mathcal{V}$ is the vocabulary,
i.e., the collection of all tokens the model can handle,
$\theta \in \mathcal{R}^{|\mathcal{V}\times d|}$ is the learnable embedding matrix.

Although simple and efficient,
this Word Embedding Averaging encoder suffers from two issues:
1) \textit{Order}.
The averaging operator discards word order information in the input sequence,
and thus poorly represents variable names where this order is important,
e.g., \lt{idx\_to\_word} and \lt{word\_to\_idx}.
2) \textit{Token importance}.
An unweighted average of word embeddings ignores the relative importance of
words in a variable name, as well as the fact that
the importance of a word can vary by context.

\begin{figure*}
    \centering
    \includegraphics[width=0.95\textwidth]{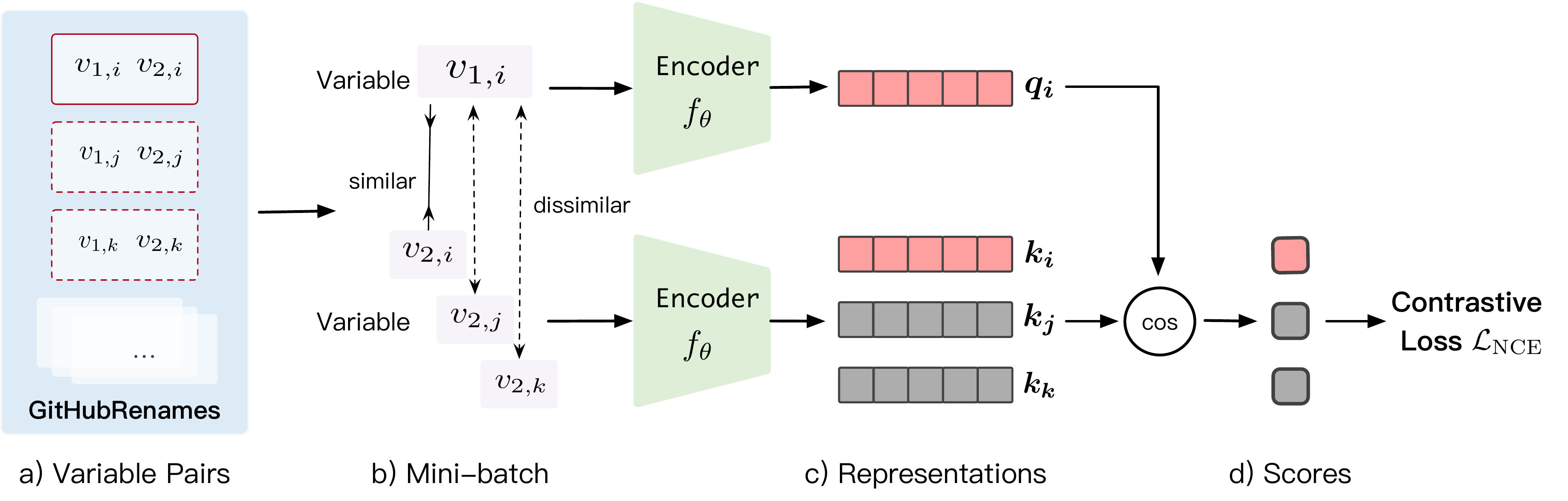}
    \caption{\small
        Overview of \modelname's contrastive pre-training method.
        a) \dataname contains interchangeable variable pairs.
        b) At each training step, sample a mini-batch of variable pairs,
        and aim to pull close the variables representations within a pair,
        e.g., $v_{1, i}$ and $v_{2, i}$,
        while pushing away the representations of other variables,
        e.g., $v_{1, i}$ and $v_{2, j}$.
        c) To achieve this, an encoder $f_\theta$ with learnable parameters $\theta$,
        is adopted to encode the variable string to hidden vectors.
        d) contrastive loss is calculated based on the similarity scores
        as the cosine distance between encoded hidden vectors;
        the encoder $f_\theta$ is optimized with gradient descent.
    }
    \label{fig:framework}
\end{figure*}

\vspace{-1ex}
\paragraph{LSTM Encoder}
Recurrent Neural Networks (RNNs)~\cite{rumelhart1986learning} generalize
feed-forward neural networks to sequences.
Given the tokenized variable name $v = (v_1, v_2, \ldots, v_n)$,
a standard RNN computes a sequence of hidden vectors
$(\boldsymbol{h_1}, \boldsymbol{h_2}, \ldots, \boldsymbol{h_n})$.
\begin{align}
    \boldsymbol{h_{t}} = \operatorname{sigmoid}\left(
    W^{\mathrm{hx}} L_{\theta_{e}}(v_t) + W^{\mathrm{hh}} \boldsymbol{h_{t-1}}
    \right),
\end{align}
where $W^{\mathrm{hx}}, W^{\mathrm{hh}} \in \mathcal{R}^{d\times d}$
are weight matrices,
and $\theta_e$ is the embedding matrix (as in \cref{eq:word_avg}).
RNNs process the input sequence by reading in one token $v_t$ at a time
and combining it with the past context $h_{t-1}$.
This captures sequential order information.
After processing all input tokens,
we can average the hidden states at each step to output a representation of the original variable:
\begin{align}
    \boldsymbol{h} = \frac{1}{n}\sum_{i=1}^{n}\boldsymbol{h_i}.
\end{align}

%In practice,
%RNNs suffer from problems such as gradient vanishing and explosion.
We use bi-directional Long Short-Term Memory (LSTM)
models~\cite{hochreiter1997long},
a variant of RNNs widely used in natural language processing.
LSTMs introduce several new components, including the input and forget
gates, controlling how much information flows from the current token, and how
much to keep from past contexts, respectively.
This better handles the token importance problem by dynamically controlling
the weight of the input token at each step.

% \begin{equation}
%     \begin{aligned}
%         i_{t}         & =\sigma\left(W^{(i)} x_{t}+U^{(i)} h_{t-1}\right) \\
%         f_{t}         & =\sigma\left(W^{(f)} x_{t}+U^{(f)} h_{t-1}\right) \\
%         o_{t}         & =\sigma\left(W^{(o)} x_{t}+U^{(o)} h_{t-1}\right) \\
%         \tilde{c}_{t} & =\tanh \left(W^{(c)} x_{t}+U^{(c)} h_{t-1}\right) \\
%         c_{t}         & =f_{t} \circ c_{t-1}+i_{t} \circ \tilde{c}_{t}    \\
%         h_{t}         & =o_{t} \circ \tanh \left(c_{t}\right)
%     \end{aligned}
% \end{equation}

\vspace{-1ex}
\paragraph{BERT}
Transformer-based models~\cite{transformer}
typically outperform LSTMs and are considered to be the
better architecture for many NLP tasks.
Pre-trained Language Models (PLMs), built upon Transformers,
can leverage massive amounts of unlabeled data and computational resources to
effectively tackle a wide range of natural language processing tasks.
Useful PLMs for programming languages include CodeBERT~\cite{feng2020codebert} and Codex~\cite{chen2021evaluating}
PLMs not only capture component ordering and token importance that LSTMs do,
but provide additional benefits:
1) BERT-based models are already pre-trained with self-supervised objectives such as
Masked Language Modeling (MLM) \cite{devlin2019bert} on a large amount of unlabeled data.
It provides a good initialization to the model parameters and improves the
model's generalization ability,
requiring fewer data to achieve satisfactory performance~\cite{brown2020language}.
2) Transformer encoders are much more powerful than previous models
thanks to the multi-head self-attention mechanism,
allowing for the model to be much wider and deeper with more parameters.
% We omit \qc{Shall we?} the details of BERT models.
We therefore propose to use PLMs for programs
as our most powerful choice of variable name encoder.

\paragraph{Effectiveness versus efficiency}
Although BERT has the largest model capacity of these encoders,
it also requires higher computation cost for both training and inference,
and suffers from a longer inference latency.
The trade-off posted between effectiveness and efficiency can vary
according to different downstream applications.
Therefore, we find it meaningful to compare all encoders in \modelname.
Different or better encoder models
can be directly plugged into the \modelname framework in the future.
We omit further interior technical details of both LSTM and BERT models as they are beyond the scope of this paper.

\subsection{Contrastive Learning Pre-training}
\label{sec:pretraining}

\modelname implements the design choices for input data, variable tokenization, and input
encoder in a contrastive learning framework.
\cref{fig:framework} provides an overview.
Conceptually, contrastive learning uses encoder
networks to encode instances (in this task, variables) into
representations (i.e., hidden vectors), and aims to minimize the
distance between similar instances while maximizing the distance
between dissimilar instances.

Specifically, given a choice of encoder and set of labeled ``positive pairs'' of variable names,
we use instance discrimination~\cite{wu2018unsupervised} as
our pre-training task,
and InfoNCE~\cite{oord2018representation} as our learning objective.
Given a mini-batch of encoded and L2-normalized representations of $K$ similar variable pairs
$\left\{(v_{1, i}, v_{2, i})| i = 1, \ldots, K \right\}$,
we first encode them to hidden representations:
\begin{align}
    \boldsymbol{q_i} & = \frac{f_\theta(v_{1, i})}{\left\|f_\theta(v_{1, i})\right\|_2}, \\
    \boldsymbol{k_i} & = \frac{f_\theta(v_{2, i})}{\left\|f_\theta(v_{2, i})\right\|_2},
\end{align}
where $\|\cdot\|_2$ is $\ell_2$-norm, $f_\theta$ denotes the encoder.
Then, we define the InfoNCE loss as:
\begin{equation}
    \mathcal{L}_{\mathrm{NCE}}\left(\boldsymbol{q}, \boldsymbol{k}\right)=
    -\mathbb{E}
    \left(
    \log
    \frac{
        e^{\boldsymbol{q_i}^{\top}\boldsymbol{k_i}/\tau}
    }
    {
        \sum_{j=1}^{K} e^{\boldsymbol{q_i}^{\top}\boldsymbol{k_j}/\tau}
    }
    \right),
\end{equation}
where $\tau$ is the temperature hyperparameter introduced by~\cite{wu2018unsupervised}.
Intuitively, this objective encourages the model to
discriminate the corresponding similar instance $v_{2, i}$ of an instance $v_{1, i}$
from other instances in the mini-batch $v_{2, j}$.
This learning objective is very similar to the cross-entropy loss for classification tasks,
while the difference is that instead of a fixed set of classes,
it treats each instance as a distinct class.
Following \cite{grill2020bootstrap},
we further make the loss symmetric and minimize the following objective function:
\begin{equation}
    \mathcal{L}=
    \frac{1}{2}\mathcal{L}_{\mathrm{NCE}}\left(\boldsymbol{q}, \boldsymbol{k}\right) +
    \frac{1}{2}\mathcal{L}_{\mathrm{NCE}}\left(\boldsymbol{k}, \boldsymbol{q}\right).
\end{equation}

In our task, this objective encourages the encoder to push the representations of
a pair of similar variables to be close to each other,
so that they can be discriminated from other variables.

We refer to this process as pre-training in the sense that the
training is not intended for a specific task
but is learning a general-purpose variable representation.

\section{Experiments}
\label{sec:eval}

In this section, we evaluate \modelname's ability to train models for variable
representation along several axes.  Section~\ref{sec:setup} addresses setup,
datasets, and baselines common to the experiments.  Then, we begin by addressing
a central claim: How well do \modelname models encode variable \emph{similarity}, as
distinct from \emph{relatedness}?
We answer this question by using pre-trained \modelname models to compute
similarity (and relatedness, resp) scores between pairs of variables, and
evaluate the results on human-annotated gold standard ground truth benchmark
(Section~\ref{sec:scoring}).

Next, we evaluate \modelname-trained models on two other
downstream tasks, demonstrating transferability: variable similarity
search (Section~\ref{sec:search}), and variable spelling error correction
(Section~\ref{sec:spelling}).

Finally, we conduct an ablation study (Section~\ref{sec:ablation}) looking at the influence of training data
size, pre-trained language models, and pre-trained embeddings from unsupervised
learning contribute to \modelname's effectiveness.

\subsection{Setup}
\label{sec:setup}
\paragraph{Pre-training.} For \modelavg and \modellstm, we use the
Adam optimizer~\cite{kingma2014adam} with $\beta_{1}=0.9, \beta_{2}=0.999,
    \epsilon=1 \times 10^{-8}$, a learning rate of 0.001, and early stop according
to the contrastive loss on the validation set.
We use a mini-batch size of 1024.
The input embedding and hidden representation dimensions are set to
768 and 150 respectively.
we also initialize the embedding layer with the CodeBERT pre-trained embedding layer.
For \modelbert, we use the AdamW optimizer
~\cite{loshchilov2018decoupled} with the same configuration and learning rate,
and a mini-batch size of 32.
We use the BERT model architecture~\cite{devlin2019bert} and initialize
the model with pre-trained weights from CodeBERT~\cite{feng2020codebert}.
For all three methods, we apply gradient norm clipping in the range $[-1,1]$,
and a temperature $\tau$ of $0.05$.
A summary of the hyper-parameters can be found along with our
data, code, and pre-trained models at \url{https://bit.ly/2WIalaW}.
% \jl{A pretrained \modelname model can be found at...}

\paragraph{Dataset.}
While we use the \dataname for training \modelname, we use the
\idbench~\cite{wainakh2021idbench} dataset for evaluation.\footnote{
    The \idbench \href{https://github.com/sola-st/IdBench}{evaluation scripts}
    were updated after publication, leading to minor differences in evaluation
    scores.
    We use their latest code as of May 1st, 2021 to evaluate the baselines and
    our models.}
\idbench is a benchmark specifically created for evaluating variable semantic
representations.
It contains pairs of variables assigned relatedness and similarity scores by
real-world developers.
\idbench consists of three sub-benchmarks---
\idbench-small, \idbench-medium, and \idbench-large,
containing 167, 247, 291 pairs of variables, respectively.
Ground truth scores for each pair of variable are assessed by multiple annotators.
Pairs with disagreement between annotators exceeding a particular threshold are
considered dissimilar; the three benchmarks differ in the choice of threshold.
The smaller benchmark provides samples with higher inter-annotator agreement,
while the larger benchmark provides more samples with commensurately lower agreement.
The medium benchmark strikes a balance.
We describe customizations of the \idbench dataset to particular tasks in their
respective sections.

\paragraph{Baselines.} We compare \modelname models to the previous state-of-the-art
as presented in \idbench~\cite{wainakh2021idbench}. We reuse the baseline
results provided by the \idbench framework. The \idbench paper evaluates a number of previous
approaches as well as a new ensemble method that outperforms them; we include as
baselines a subset of those previous techniques, and the ensemble
method. Of the string distance/syntactic functions (still broadly used in
various name-related applications~\cite{rice2017detecting,liu2016nomen}), we
include \textbf{Levenshtein Edit Distance (LV)} (the number of single-character
edits required to transform one string into the other); it performs in the top
half of techniques on scoring similarity, and is competitive with the other
syntactic distance metric~\cite{needleman1970general} on relatedness.
Of the embedding-based single models, we include
\textbf{FastText CBOW (FT-cbow) and SG
    (FT-sg)}~\cite{bojanowski2017enriching}, extensions of \wordtvec that
incorporate subword information, to better handle infrequent
words and new words.
These were the best-performing embedding-based methods on both relatedness and
similarity.

Finally, we include two \textbf{combined} models.
\idbench~\cite{wainakh2021idbench} proposes an ensemble method that combines the
scores of all models and variable features. For each pair in \idbench, the
combined model trains a Support Vector Machine (SVM) classifier with all other
pairs, then applies the trained model to predict the score of the left-out pair.
Note that this approach is trained on the \idbench benchmark itself and is not
directly comparable to other methods. For comparison, we add \modelavg,
\modellstm, \modelbert scores as additional input features to the
combined approach, and report the results for Combined-\modelname.

% \begin{itemize}
%     \item \textbf{Word2vec Continuous Bag of Words (w2v-cbow) and Continuous
%       Skip-Gram (w2v-sg)}, two variants of Word2vec
%       objectives. 
%     \item \textbf{Path-based}~\cite{alon2018general}, which treats programs
%       as Abstract Syntax Trees (ASTs) instead of flat sequences of tokens before
%       applying learning algorithms such as Word2vec, allowing the model to
%       leverage the structured nature of code.
%\end{itemize}

\subsection{Variable Similarity and Relatedness Scoring}
\label{sec:scoring}

Our central claim is that \modelname is well-suited to capturing and predicting variable
similarity.
Formally, given two variables $u$ and $v$, we obtain variable representations
with pre-trained \modelname encoder $f_{\theta'}$ and
compute the variable similarity score as the cosine similarity between the two vectors:
\begin{align}
    \boldsymbol{h_u},\boldsymbol{h_v} & = f_{\theta'}(u), f_{\theta'}(v) \\
    \hat{s}(u, v)                     & =
    \frac{
        \boldsymbol{h_u} \cdot \boldsymbol{h_v}
    }{
        \|\boldsymbol{h_u}\|_2\|\boldsymbol{h_v}\|_2
    },
\end{align}
where $\hat{s}(u, v)$ denotes the \modelname's predicted similarity score.
Following \idbench~\cite{wainakh2021idbench}, we then compare the similarity
scores of pre-trained \modelname representations with human ground-truth
similarity scores by computing Spearman's rank correlation coefficient
between them.
% With a benchmark dataset $\mathcal{B} = {(u_i, v_i, s_i)}$ containing variable
% pairs $u_i$, $v_i$ and a human-assigned ground-truth similarity score $s_i$, the
% Spearman's rank correlation coefficient is computed as:
%  \begin{align}
%      S & = \operatorname{spearmanr}
%      \{(\hat{s}(u_i, v_i), s_i)\}_{(u_i, v_i, s_i)\in \mathcal{B}}.
%  \end{align}
This correlation coefficient falls in the range [-1, 1], where 1 indicates
perfect agreement between the rankings; -1 indicates perfect disagreement; and 0
indicates no relationship.

Note that the \modelname pre-training task is explicitly optimizing the distance
between similar variable pairs. Thus, the variable similarity scoring task only
really evaluates the performance of the pre-training itself. To more fully evaluate whether
our method leads to better representations that can transfer, we also evaluate
on the variable relatedness scoring task.

\paragraph{Results}
\cref{tbl:spearman} shows the models' performance on the
similarity and relatedness tasks in terms of Spearman's rank
correlation with ground truth.
\cref{tbl:idbench_score} shows that \modelbert improves over the previous state-of-the-art
on all three \idbench benchmarks,
with an absolute improvement of 0.18 on \idbench-small and 0.13 on \idbench-large
compared to the previous best approach, FT-cbow.
This shows that \modelname aligns
much better with human developers' assessment of variable similarity than any of
the previously proposed models.
Interestingly, \modelavg also outperforms FT-cbow by a large margin (+0.12 on \idbench-small).
This suggests that most of our gains do not come from the use of a more powerful encoder
architecture such as BERT.
Instead, we conclude that the \dataname dataset is effective at providing supervision signals of
variable similarity, and the contrastive learning objective is effective.
Although their architectures are very similar, \modelavg
outperforms FT-cbow.

That said, the improvements in \modelbert (+0.06) and \modelname (+0.03) over
\modelavg verify our assumption that powerful models with larger
representational capacity are necessary for learning better variable
representations, since they are able to capture and encode more information
(e.g., sequential order and token importance) than the embedding averaging
methods.

\cref{tbl:idbench_score_relatedness} shows that
\modelname also achieves the state-of-the-art performance on \idbench in terms of relatedness prediction.
It surpasses the previous best by 0.07 on \idbench-small and
0.07 on \idbench-large.
This is noteworthy because \modelname training does not explicitly optimize
for relatedness.
This suggests that the \modelname pre-training task learns better generic
representations, rather than overfitting to the target task (i.e., variable
similarity).
This is very important, and supports our major contribution: By pre-training for
the similarity learning task on \dataname with a contrastive objective,
\modelname achieves better representations which can be applied to general
tasks.

\begin{table}
    \caption{Spearman's rank correlation with \idbench-small,
        \idbench-medium, \idbench-large of single models (top) and ensemble models
        (bottom), by increasing performance.}
    \label{tbl:spearman}
    \centering
    \subfloat[Similarity scores]{    \label{tbl:idbench_score}\begin{tabular}{lrrrr}
            \toprule
            Method              & Small         & Medium        & Large         \\
            \midrule
            %        w2v-cbow            & 0.11          & 0.11          & 0.10          \\
            %        w2v-SG              & 0.15          & 0.18          & 0.15          \\
            %        Path-based          & 0.23          & 0.22          & 0.21          \\
            %        NW                  & 0.30          & 0.27          & 0.27          \\
            FT-SG               & 0.30          & 0.29          & 0.28          \\
            LV                  & 0.32          & 0.30          & 0.30          \\
            FT-cbow             & 0.35          & 0.38          & 0.38          \\
            \modelavg           & 0.47          & 0.45          & 0.44          \\
            \modellstm          & 0.50          & 0.49          & 0.49          \\
            \modelbert          & \textbf{0.53} & \textbf{0.53} & \textbf{0.51} \\
            \midrule
            Combined-\idbench   & 0.48          & 0.59          & 0.57          \\
            Combined-\modelname & \textbf{0.66} & \textbf{0.65} & \textbf{0.62} \\
            \bottomrule
        \end{tabular}}

    \subfloat[Relatedness scores]{\label{tbl:idbench_score_relatedness}\begin{tabular}{lrrrr}
            \toprule
            Method              & Small         & Medium        & Large         \\
            \midrule
            %       w2v-cbow            & 0.41          & 0.42          & 0.38          \\
            LV                  & 0.48          & 0.47          & 0.48          \\
            %       NW                  & 0.49          & 0.47          & 0.46          \\
            %       Path-based          & 0.53          & 0.58          & 0.60          \\
            %       w2v-SG              & 0.59          & 0.62          & 0.60          \\
            FT-SG               & 0.70          & 0.71          & 0.68          \\
            FT-cbow             & 0.72          & 0.74          & 0.73          \\
            \modelavg           & 0.67          & 0.66          & 0.66          \\
            \modellstm          & 0.71          & 0.70          & 0.69          \\
            \modelbert          & \textbf{0.79} & \textbf{0.79} & \textbf{0.80} \\
            \midrule
            Combined-\idbench   & 0.71          & 0.78          & 0.79          \\
            Combined-\modelname & \textbf{0.79} & \textbf{0.81} & \textbf{0.85} \\
            \bottomrule
        \end{tabular}}
    \vspace{-0.1in}
\end{table}

\subsection{Variable Similarity Search}
\label{sec:search}

\begin{figure}[t]
    \centering
    \subfloat[Similarity Search]{
        \label{fig:sim_search}
        \includegraphics[width=0.48\linewidth]{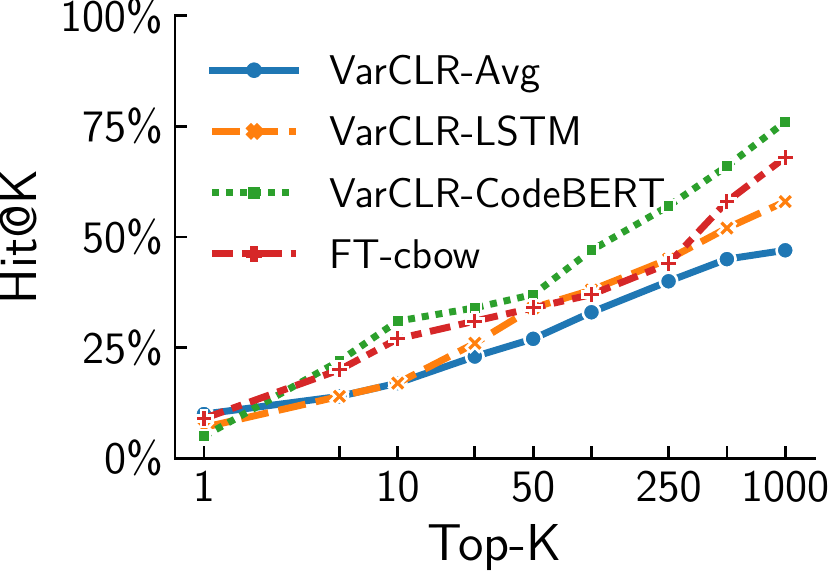}
    }
    \subfloat[Spelling Error Correction]{
        \label{fig:spelling_corr}
        \includegraphics[width=0.48\linewidth]{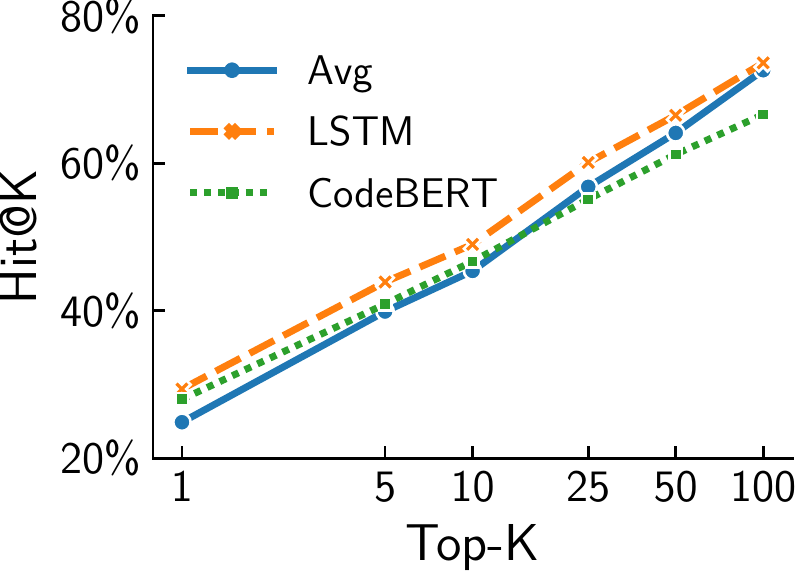}
    }
    \vspace{-0.1in}
    \caption{Hit@K score comparison on \idsim and \vartypo.}
    \vspace{-0.2in}
\end{figure}

\begin{table*}
    \caption{\small Variable Similarity Search.
        Top-5 most similar variables found by the \idbench method and \modelbert.}
    \label{tbl:similarity_search}
    \footnotesize
    \centering
    \begin{tabular}{lllllll}
        \toprule
        Variable                          & Method     & \multicolumn{5}{c}{Top 5 Similar Variables}                                                                                             \\
        \midrule
        \multirow{2}{*}{\tt{substr}}      & FT-cbow    & \tt{substring    }                          & \tt{substrs           } & \tt{subst        }   & \tt{substring1   }    & \tt{substrCount} \\
                                          & \modelbert & \tt{subStr    }                             & \tt{substring}          & \tt{substrs}         & \tt{stringSubstr}     & \tt{substrCount} \\
        \midrule
        \multirow{2}{*}{\tt{item}}        & FT-cbow    & \tt{itemNr       }                          & \tt{itemJ             } & \tt{itemL        }   & \tt{itemI        }    & \tt{itemAt}      \\
                                          & \modelbert & \tt{pItem}                                  & \tt{itemEl}             & \tt{mItem}           & \tt{itemEls}          & \tt{itemValue}   \\
        %        \midrule
        %        \multirow{2}{*}{\tt{count}}       & FT-cbow             & \tt{countTbl     }                          & \tt{countInt          } & \tt{countRTO     }   & \tt{countsAsNum  }    & \tt{countOne}    \\
        %                                          & \modelbert & \tt{sCount}                                 & \tt{countOf}            & \tt{counts}          & \tt{countInt}         & \tt{countTh}     \\
        %        \midrule
        %        \multirow{2}{*}{\tt{rows}}        & FT-cbow             & \tt{rowOrRows    }                          & \tt{rowXs             } & \tt{rows\_1      }   & \tt{rowsAr       }    & \tt{rowIDs}      \\
        %                                          & \modelbert & \tt{drows}                                  & \tt{allrows}            & \tt{rowsArray}       & \tt{ows}              & \tt{nRows}       \\
        \midrule
        \multirow{2}{*}{\tt{setInterval}} & FT-cbow    & \tt{resetInterval}                          & \tt{setTimeoutInterval} & \tt{clearInterval}   & \tt{getInterval  }    & \tt{retInterval} \\
                                          & \modelbert & \tt{pInterval}                              & \tt{mfpSetInterval}     & \tt{setTickInterval} & \tt{clockSetInterval} & \tt{iInterval}   \\
        \midrule
        \multirow{2}{*}{\tt{minText}}     & FT-cbow    & \tt{maxText      }                          & \tt{minLengthText     } & \tt{microsecText }   & \tt{maxLengthText}    & \tt{minuteText}  \\
                                          & \modelbert & \tt{minLengthText}                          & \tt{minContent}         & \tt{maxText}         & \tt{minEl}            & \tt{min}         \\
        \midrule
        \multirow{2}{*}{\tt{files}}       & FT-cbow    & \tt{filesObjs    }                          & \tt{filesGen          } & \tt{fileSets     }   & \tt{extFiles     }    & \tt{libFiles}    \\
                                          & \modelbert & \tt{filesArray}                             & \tt{aFiles}             & \tt{allFiles}        & \tt{fileslist}        & \tt{filelist}    \\
        \midrule
        \multirow{2}{*}{\tt{miny}}        & FT-cbow    & \tt{min\_y       }                          & \tt{minBy             } & \tt{minx         }   & \tt{minPt        }    & \tt{min\_z}      \\
                                          & \modelbert & \tt{ymin}                                   & \tt{yMin}               & \tt{minY}            & \tt{minYs}            & \tt{minXy}       \\
        \bottomrule
    \end{tabular}
\end{table*}

We next evaluate our learned representations in the context of a more applied
downstream application: similar variable search.
Similar variable search identifies similar variable names in
a set of names given an input query.
This can be useful for refactoring code, or for assigning variables more
readable names (e.g., replacing \lt{fd} with \lt{file_descriptor}).
For a given set of variables $\mathcal{V}$ and a pre-trained \modelname encoder $f_{\theta'}$,
we compute representation vectors $\mathcal{K} = \{f_{\theta'}(v) | v\in\mathcal{V}\}$.
For a query variable $u$, we find top-$k$ similar variables in $\mathcal{V}$
with the highest cosine similarity to $f_{\theta'}(u)$.

To quantitatively evaluate effectiveness in finding similar variables,
we created a new mini-benchmark \idsim from  the original \idbench benchmark.
We select variable pairs which have human-assessed similarity scores
greater than 0.4 in \idbench.
This leaves us with 100 `similar' variable pairs from
all 291 variable pairs in the \idbench-large benchmark.
We use the variable collection provided in \idbench containing 208,434 variables
as the overall candidate pool.
We use Hit@K as our evaluation metric, computing
the cosine similarity of the representations
of a query variable $u$ and all the variables in the candidate pool.
We select the top-K variables with the highest similarity scores and check whether
the corresponding similar variable $v$ is in the top-K list.
We choose K to be 1, 5, 10, 25, 50, 100, 250, 500, 1000.

\paragraph{Results}
As shown in \cref{fig:sim_search},
\modelbert achieves the best similarity search performance,
with 47\% at K=100 and 76\% at K=1000,
compared to FT-cbow (37\% at K=100, 68\% at K=1000).
This indicates that our method is effective at finding similar variables,
able to distinguish the most similar variable to the query variable out of 200 distractors
around 76\% of the time.\footnote{
    Since our Hit@1000 score in a candidate pool of size $\sim$200,000 is 76\%.
    Although inspecting the top 1000 may not be practical as an real-world application itself,
    it is still an informative metric of the representation quality, 
    and may indicate effectiveness in other settings,
    e.g., a developer looking at the top 5 similar variables from 1,000 candidates,
    or using \modelname as generate a large candidate pool as the first stage.
    }
Interestingly, \modelavg and \modellstm are less effective at
similarity search than FT-cbow,
even though they outperform FT-cbow by a large margin in the similarity scoring task.
Embedding-based methods are still a strong baseline
for variable similarity search.  However,
contrastive methods still amplify the effectiveness of unsupervised embedding methods.

Similarity scoring and similarity search are distinct tasks, and so
it is not unexpected that techniques will be equally effective on both.
For example, \wordtvec tends to put the embeddings of similar rare words close to some
common frequent word.
This behavior does not affect the similarity search effectiveness because the rare words
are able to find each other, and the frequent word is close enough to \textit{its}
similar word than to these rare words.
However, this will hurt similarity scoring between the rare words and the frequent variable,
since they are actually not similar.
In comparison, \modelname is able to avoid these kinds of scoring mistakes.
%since the contrastive learning objective automatically pushes away the representations
%of dissimilar pairs.

\paragraph{Case Study.}
We demonstrate our results qualitatively by choosing the same set of variables
used to demonstrate this task in the \idbench paper,
and displaying the comparative results in \cref{fig:sim_search}.
For space, we omit two of the variables (\lt{rows} and \lt{count}) in the set;
the two methods perform comparably (such as on \lt{substr}).
We observe that the overall qualities of the two methods' results
are similar.
This is understandable since the gap between the two methods on variable similarity
search is relatively small as shown in \cref{tbl:similarity_search}.

Meanwhile, it is worth noting that \modelbert is better at penalizing
distractive candidates that are only related but not similar.
For example, for \lt{minText},
\modelbert ranks \lt{minLengthText}, \lt{minContent} before \lt{maxText},
while FT-cbow suggests the opposite.
For \lt{miny}, \modelbert ranks \lt{ymin}, \lt{yMin}, \lt{minY} as top-3,
while FT-cbow suggests related but dissimilar variables such as \lt{minBy} and \lt{minx}.
This provides additional evidence that our method is able to better represent
semantic similarity rather than pure relatedness.

\subsection{Variable Spelling Error Correction}
\label{sec:spelling}

Spelling Error Correction is a fundamental yet challenging task
in natural language processing~\cite{jayanthi2020neuspell}.
We explore the possibility of applying \modelname models to perform
spelling error correction on variable names.
If the representations of misspelled variable names are close to their correct
versions, corrections may be found via nearest neighbor search.
Fortunately, the \dataname dataset enables this goal, because
a portion of renaming edits in \dataname
are actually correcting spelling errors in previous commits.
We can therefore reformulate this problem as a variable similarity search task,
since our method treats these misspelled names as similar to their corrected
versions.

We create a new synthetic variable spelling error correction dataset, \vartypo,
with 1023 misspelled variables and their corrections.
Specifically, we create this dataset by sampling variables from the 208,434 variable pool from \idbench,
and use the \texttt{nlpaug}\footnote{\url{https://github.com/makcedward/nlpaug}} package~\cite{ma2019nlpaug}
to create misspelled variables from the correct ones.
We use \texttt{KeyboardAug} which simulates typo error according to
characters' distance on the keyboard.
This task is challenging
because our method does not leverage any external dictionary
or hand-crafted spelling rules.
Meanwhile, although string distance functions such as Levenshtein distance
can potentially perform better, these functions require expensive one-by-one comparisons
between the query variable and every variable in the pool, which is very time consuming,
while our method uses GPU-accelerated matrix multiplication to compute all cosine distances at once
and can potentially adopt an even more efficient vector similarity search library such as \texttt{faiss}.
Therefore,
we believe it is still an informative benchmark for evaluating variable representations.

\paragraph{Results}
Similar to variable similarity search,
we evaluate the effectiveness as the Hit@K score of
using the representation of misspelled variables
to retrieve the corresponding correct variable.
As shown in \cref{fig:spelling_corr},
\modelname can successfully correct the 29.4\% of the time at Top-1,
and 73.6\% of the time at Top-100.
One interesting observation we find is that in this task,
the gap (-4.5\% at Top-1 and -1.0\% at Top-100) between \modelavg and
the other two powerful encoders is relatively small.
It even outperforms \modelbert after K=25.
One possible explanation is that fixing a typo requires neither
word sequential order or word importance information,
i.e., being able to model the variable as a sequence instead of a bag of words does not
benefit this task.

\paragraph{Case Study.}
For illustration, we randomly select misspelled variable names and
use our \modelname to find the most similar correct variable names.
As shown in \cref{tbl:spelling_corr}.,
our model is able to correct some of the misspelled variables, including
insertions, deletions, and modifications, while failing to recover others.
Notably, variable names consisting of multiple words such as \lt{minSimilarity}
can be corrected successfully.

% https://github.com/client9/misspell/blob/master/words.go

\begin{table}
    \caption{\small
        The top-3 most similar variables to misspelled variables, found by \modelname.}
    \label{tbl:spelling_corr}
    \centering
    \footnotesize
    \begin{tabular}{llll}
        \toprule
        Variable                  & Top 3 Similar Variables                                 \\
        \midrule
        \tt{tem\red{ep}ratures}   & \tt{temperatures}, \tt{temps}, \tt{temlp}               \\
        %       \tt{maxTem\red{ep}ratures} & \tt{maxTemperatures}, \tt{maxTextures}, \tt{maxTolerance} \\
        \tt{similar\red{l}ity}    & \tt{similarity}, \tt{similarities}, \tt{similar}        \\
        \tt{minSimilar\red{l}ity} & \tt{minSimilarity}, \tt{similarity}, \tt{minRatio}      \\
        \tt{program\red{\_}able}  & \tt{programmable}, \tt{program}, \tt{program6}          \\
        %       \tt{transforme\red{\_}s}   & \tt{transformations}, \tt{transformers}, \tt{transforms}  \\
        \midrule
        \tt{supervis\red{i}or}    & \tt{superior}, \tt{superview}, \tt{superc}              \\
        \tt{produc\red{it}ons}    & \tt{obligations}, \tt{proportions}, \tt{omegastructors} \\
        \tt{trans\red{al}tion}    & \tt{transac}, \tt{trans}, \tt{transit}                  \\
        \bottomrule
    \end{tabular}
\end{table}

\subsection{Ablation Studies}
\label{sec:ablation}

So far we have demonstrated the importance of both contrastive learning and
sophisticated models like CodeBERT for \modelname performance.
Here, perform ablation studies to measure the effect of additional design decisions in \modelname: of training
data size, of using pre-trained language models, and of using pre-trained
embeddings from unsupervised learning.

\subsubsection{Effect of Data Size on Contrastive Pre-training}

%We examine the impact of the data size during contrastive pre-training on the quality of the learned \modelname representations.
Pre-training \modelname requires weakly-supervised data scraped from public
repositories.
Thus, we evaluate how much data is required to train an
effective model, to elucidate data collection costs.
To evaluate this, we train \modelavg, \modellstm, \modelbert
on 0\%, 0.1\%, 1\%, 3.16\%, 10\%, 21.5\%, 46.4\%, 100\% percent of
the full dataset, measuring the similarity score on \idbench-medium.

\begin{figure}
    \centering
    \includegraphics[width=0.70\columnwidth]{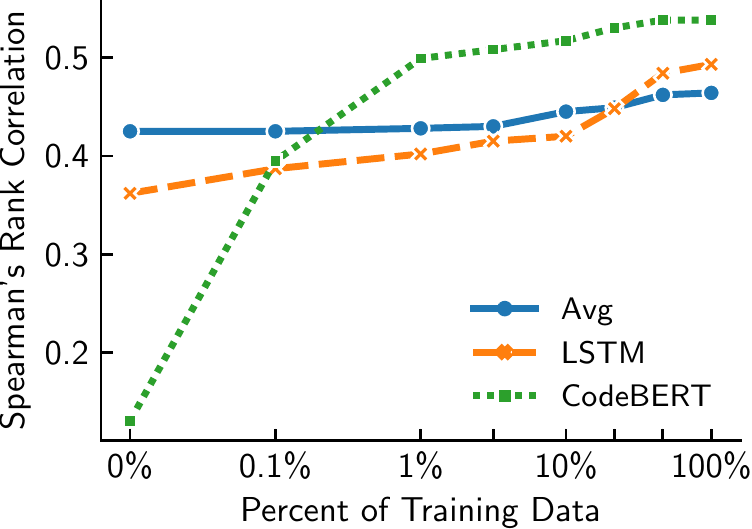}
    \caption{\small Effect of contrastive pre-training data size on learned \modelname representations, evaluated on \idbench-medium.}
    \label{fig:data_size}
\end{figure}

\cref{fig:data_size} shows the results.
For all three \modelname variants,
training data size has a significant positive effect on effectiveness.
This is especially true for \modelbert, but
performance flattens and converges as training data size approaches 100\%.
This suggests that \dataname is of an appropriate size for this task.

Another interesting observation is that \modelavg outperforms
\modellstm with smaller amounts of training data.
This indicates the more powerful LSTM model does not
surpass a simple one until the data size reaches a critical threshold.
This is likely because a more complex model has more parameters to train and
requires more data to reach convergence.
With sufficient data, larger models win, thanks to their
representational capacity.
This suggests a caveat in applying representation learning models:
it is important to choose a model with an appropriate complexity given the
amount of available data, rather than defaulting to the best-performing model
overall.

\subsubsection{Using a Pre-trained Language Model}

Before contrastive pre-training on \dataname, \modelbert is initialized with a model
(pre-)pre-trained on a large code corpus.
The effect of this pre-training is also illustrated in \cref{fig:data_size}.
Although \modelbert has a much larger number of parameters,
it outperforms \modelavg and \modellstm after contrastive pre-training on only
1\% of \dataname.
While this seems to contradict the conclusion reached in the comparison between
\modellstm and \modelavg, it displays the benefit of initialization with a
pre-trained model.
Compared to \modellstm, which contains randomly initialized parameters that have
to be trained from scratch, \modelbert parameters produce reasonable
representations from the start.
Therefore, it requires less data to converge, and thanks to its large model
capacity, ultimately outperforms the other two variants by a large
margin.

Despite the fast convergence, directly applying CodeBERT without contrastive
pre-training leads to poor performance (0.13 at 0\% data).
One possible reason is that CodeBERT was originally trained for whole-program
representations, and using it with variable names as inputs leads to a
problematic divergence
from its training data distribution.

\subsubsection{Effect of Pre-trained CodeBERT Embeddings}

\begin{table}
    \caption{\small Effect of pre-trained CodeBERT embeddings on similarity score
        effectiveness (Spearman's). Models are either randomly
        initialized and contrastively pre-trained (Contrastive), initialized with
        CodeBERT embeddings (CodeBERT), or both (\modelname).  }
    \label{tbl:effect_emb}
    \centering
    \begin{tabular}{lrrrr}
        \toprule
        Method                   & Small         & Medium        & Large         \\
        \midrule
        Contrastive-\textsc{Avg} & 0.34          & 0.33          & 0.30          \\
        \textsc{CodeBERT-Avg}    & 0.44          & 0.43          & 0.40          \\
        \modelavg                & \textbf{0.47} & \textbf{0.45} & \textbf{0.44} \\
        \midrule
        Contrastive-LSTM         & 0.35          & 0.33          & 0.30          \\
        CodeBERT-LSTM            & 0.36          & 0.36          & 0.36          \\
        \modellstm               & \textbf{0.50} & \textbf{0.49} & \textbf{0.49} \\
        \bottomrule
    \end{tabular}
\end{table}

Both \modelavg and \modellstm are initialized with the word embeddings from
CodeBERT before contrastive pre-training.
To study the effect of these pre-trained embeddings, we measure the Spearman's
correlation coefficient of the similarity scores of the models modified in two
ways: one with randomly-initialized embeddings that is then contrastively
pre-trained (``Contrastive'' in \cref{tbl:effect_emb}), and one that is
initialized with CodeBERT embeddings but \emph{not} contrastively pre-trained
(``CodeBERT'' in \cref{tbl:effect_emb}).

The results show that pre-trained CodeBERT embeddings are essential to the
performance of \modelavg and \modellstm.
However, directly adopting the pre-trained embeddings alone is still insufficient,
especially for LSTMs.
This implies that both unsupervised pre-training and weakly supervised pre-training
are indispensable for useful variable representations:
the former takes advantage of \textit{data quantity} by leveraging a huge amount of
unlabeled data, while the latter takes advantage of \textit{data quality} using
the weakly supervised \dataname dataset.

\iffalse
    % FIXME: PUT THIS IN THE DATA?CODE RELEASE
    \begin{table}[t]
        \centering
        \caption{Summary of \modelname encoder hyperparameters.}
        \begin{tabular}{lrrr}
            \toprule
            Hyperparameter     & Avg       & LSTM      & BERT      \\
            \midrule
            % contrastive learning
            Batch size         & 1024      & 1024      & 64        \\
            Temperature $\tau$ & 0.05      & 0.05      & 0.05      \\
            % training
            Training epochs    & 30        & 16        & 1         \\
            Patience           & 10        & 10        & -         \\
            Learning rate      & $10^{-4}$ & $10^{-4}$ & $10^{-4}$ \\
            % architecture
            Embedding dim      & 768       & 768       & 768       \\
            Hidden dim         & -         & 150       & 768       \\
            Intermediate dim   & -         & -         & 3072      \\
            Number of layers   & -         & 1         & 12        \\
            Dropout rate       & 0.5       & 0.5       & 0.1       \\
            % optimization
            Adam $\epsilon$    & $10^{-8}$ & $10^{-8}$ & $10^{-8}$ \\
            Adam $\beta_1$     & 0.9       & 0.9       & 0.9       \\
            Adam $\beta_2$     & 0.999     & 0.999     & 0.999     \\
            Gradient clipping  & 1.0       & 1.0       & 1.0       \\
            \bottomrule
        \end{tabular}
        \label{tbl:hyper}
    \end{table}
\fi

\section{Related Work}

\paragraph{Variable Names and Representations}
Variable names are important for source code readability and
comprehension~\cite{Gellenbeck1991,Lawrie2006}.
Because of this, there has been recent work focusing on automatically suggesting
clear, meaningful variable names for tasks such as code
refactoring~\cite{Allamanis2014,Context2Name,liu2019learning} and
reverse engineering~\cite{Jaffe2018,lacomis2019dire}.

A common approach involves building prediction engines on top of learned
variable representations.
Representation learning is a common task in Natural Language Processing (NLP),
and these techniques are often adapted to source code.
Simpler approaches model variable representations by applying
\wordtvec~\cite{word2vec} to code
tokens~\cite{mikolov2013distributed,bojanowski2017enriching,wainakh2021idbench,chen2019literature},
while more advanced techniques have adapted neural network
architectures~\cite{Katz2018} or pre-trained language
models~\cite{chen2021evaluating}.
Source code representation is a common enough task that researchers have developed
benchmarks specifically for variable~\cite{wainakh2021idbench}
and program representations~\cite{wang2019coset}.

\paragraph{Similarity and Relatedness}
A fundamental concern with existing variable representations and suggestion engines is the
difference between ``related'' and ``similar''
variables~\cite{miller1991contextual,wainakh2021idbench}.
``Related'' variables reference similar core concepts without concern for their
precise meaning, while ``similar'' variables are directly interchangeable.
For example, \lt{minWeight} and \lt{maxWeight} are related but not similar,
while \lt{avg} and \lt{mean} are both.
Unlike state-of-the-art techniques, which only model relatedness, \modelname
explicitly optimizes for similarity by adapting contrastive learning techniques from NLP and
computer vision research.
In NLP, systems are often designed to focus on text
relatedness~\cite{finkelstein2001placing,bruni2014multimodal,zesch2008using},
similarity~\cite{hill2015simlex}, or both~\cite{agirre2009study}. While document
search might only be concerned with relatedness~\cite{finkelstein2001placing}
similarity is particularly important in systems designed for paraphrasing
documents~\cite{wieting2015towards,wieting2021paraphrastic}.

\modelname relies on \emph{contrastive learning} to optimize for similarity.
Contrastive learning is particularly useful for learning visual representations
without any supervision data~\cite{wu2018unsupervised, tian2020contrastive,
    he2020momentum, chen2020simple, caron2020unsupervised, grill2020bootstrap,
    chen2021exploring}, but has also been used for NLP~\cite{mnih2013learning}.
Recent work has applied contrastive learning to the pre-training of language
models to learn text representations~\cite{clark2019electra} and, similar to our
task, learn sentence embeddings for textual similarity
tasks~\cite{gao2021simcse}.
Contrastive learning has also been used for code representation learning~\cite{jain2020contrastive} 
where source-to-source compiler transformation is applied for 
generating different views of a same program.
Different from this work, we focus on learning representations for variable names,
and leverage additional data from GitHub for better supervision.

\paragraph{String similarity and spelling errors.}
Efficient string similarity search remains an active research area
\cite{li2008efficient,zhang2010bed,deng2013top,bayardo2007scaling}.
Most of these methods can be categorized as \textit{sparse retrieval} methods,
focusing on distance functions on the original string or n-grams.
These algorithms depend on the lexical overlap between strings
and thus cannot capture the similarity between variables pairs such as
\lt{avg} and \lt{mean}.
More recently, \textit{dense retrieval} methods have been shown effective
in NLP tasks~\cite{lee2019latent,karpukhin2020dense}.
These methods perform similarity search in the space of learned representations,
so that sequences with similar meanings but low lexical overlap can be found.
Meanwhile, extremely efficient similarity search frameworks
for dense vectors such as \texttt{faiss} \cite{johnson2019billion} can be applied.
\modelname introduces the concept of dense retrieval into the variable names domain,
enabling more effective and efficient finding of a list of candidates
that are similar to a given variable name.
% \clg{one more sentence putting our work in context.  Are we better?
%     Complementary?}

Neural models for spelling error correction usually require parallel training data
which are hard to acquire in practice~\cite{hagiwara2020github, zhang2020spelling}.
Recent work adopts different mechanisms to create synthetic parallel data,
including noise injection~\cite{jayanthi2020neuspell},
and back-translation models~\cite{guo2019spelling}.
We leave a detailed comparison to future work, but note that \modelname shows
promise without expensive training data.

\paragraph{Name- and Machine Learning-based Program Analyses}
Our downstream tasks are examples of program analyses based on information
gathered with machine learning (ML).
Name-based based program analyses predicated on machine learning have been used in many contexts.
In the context of code modification, they have been used for variable name suggestion
from code contexts \cite{Context2Name}, method and class name rewriting
\cite{liu2019learning} and generation \cite{allamanis2015suggesting}, code
generation directly from docstrings \cite{chen2021evaluating}, and automated
program repair~\cite{poshyvanyk2018repair,monperrus2018repair}.
They have also been used for type inference from natural language
information~\cite{xu2016python,malik2019nl2type}, detecting
bugs~\cite{pradel2011detecting,pradel2018deepbugs,rice2017detecting,kate2018phys},
and detecting vulnerabilities~\cite{harer2018automated}.
\modelname can serve as a drop-in pre-training step for such techniques,
enabling more effective use of the semantic information contained in variable
names for a wide range of such analyses.

\section{Discussion}

In this paper, we study variable representation learning, a problem with
significant implications for machine learning and name-based program analyses.
We present a novel method based on contrastive learning
for pre-training variable representations.
With our new weakly-supervised \dataname dataset, our method
enables the use of stronger encoder architectures
in place of {\tt{word2vec}}-based methods for this task,
leading to better generalized representations.
Our experiments show that \modelname
greatly improves representation quality not only in terms of variable similarity,
but also for other downstream tasks.
While these downstream tasks may not be immediately practical themselves,
our approach is promising as a drop-in pre-training solution for other variable
name-based analysis tasks, which we hope others will attempt in future work.
For example, \modelname can replace the
the word2vec-CBOW embeddings used in a name-based bug detector~\cite{pradel2011detecting},
or the n-gram based language model used as a similarity scoring function
for name suggestion~\cite{Allamanis2014}.
Existing dictionary-based IDE spell-checkers may also benefit from 
using \modelname to rank suggestions.

We note limitations and possible threats in our study. Our dataset is
automatically constructed from git commits from GitHub, and likely contains noise
that can harm contrastive learning performance~\cite{li2021align}.  However, our
results show that despite this noise, our models transfer well, and our
evaluation is based on an entirely distinct test set.  Knowledge distillation
and self-training methods~\cite{hinton2015distilling, du2021self} such as
momentum distillation~\cite{li2021align} can be applied to deal with the noise
in weak supervision data~\cite{tarvainen2017mean, li2021align}.

In this work, we applied \modelname exclusively to unsupervised downstream tasks.
Fine-tuning \modelname models is likely to enable significant performance
improvements for more complicated tasks, like natural variable name
suggestion~\cite{Allamanis2014}. Beyond constructing similar variable names, it
is also conceptually possible to construct similar pairs of larger code snippets
from git diffs describing patches. Applying contrastive learning on these pairs
can potentially improve CodeBERT code representation and understanding, which
could benefit tasks well beyond variable similarity, such as code search.
Finally, we used instance discrimination~\cite{wu2018unsupervised} to guide our
contrastive learning approach, with promising results. This suggests that more
advanced contrastive learning methods such as MoCo~\cite{he2020momentum},
BYOL~\cite{grill2020bootstrap}, SwAV~\cite{caron2020unsupervised} be adapted to
this task for better representation learning in general.

\section*{Acknowledgments}

The authors would like to thank Michael Pradel and the authors of \idbench for
providing us with data for our experiments. 
This material is based upon work supported in part by the 
National Science Foundation (awards 1815287 and 1910067).

\bibliographystyle{ACM-Reference-Format}
\bibliography{references}

\end{document}